\documentclass[10pt,letterpaper]{article}
\usepackage[top=0.85in,footskip=0.75in]{geometry}

\usepackage{changepage}

\usepackage[utf8]{inputenc}

\usepackage{textcomp,marvosym}

\usepackage{fixltx2e}

\usepackage{amsmath,amssymb}

\usepackage{cite}

\usepackage{nameref,hyperref}

\usepackage{microtype}
\DisableLigatures[f]{encoding = *, family = * }

\usepackage{rotating}

\raggedright
\usepackage[aboveskip=1pt,labelfont=bf,labelsep=period,justification=justified,singlelinecheck=off]{caption}

\makeatletter
\renewcommand{\@biblabel}[1]{\quad#1.}
\makeatother

\date{}

\usepackage{lastpage,fancyhdr,graphicx}
\usepackage{epstopdf}
\usepackage{ragged2e}
\fancyheadoffset[L]{2.25in}
\fancyfootoffset[L]{2.25in}
\usepackage{color}
\definecolor{burntorange}{RGB}{191,87,0}

\begin{document}
\vspace*{0.35in}

% Please capitalize all terms in the title except conjunctions, prepositions, and articles.
\begin{flushleft}
{\Large
\textbf\newline{Quantifying discrepancies in opinion spectra from online and offline networks}
}
\newline
\\
Deokjae Lee\textsuperscript{1},
Kyu S. Hahn\textsuperscript{2},
Soon-Hyung Yook\textsuperscript{3},
Juyong Park\textsuperscript{4,*}
\\
\bigskip
\bf{1} Center for Complex Systems Studies and CTP,  Department of Physics and Astronomy, Seoul National University, Korea
\\
\bf{2} Department of Communication, Seoul National University, Korea
\\
\bf{3} Department of Physics, Kyung Hee University, Korea
\\
\bf{4} Graduate School of Culture Technology, Korea Advanced Institute of Science \& Technology, Korea
\\
\bigskip

% Insert additional author notes using the symbols described below. Insert symbol callouts after author names as necessary.
% 
% Remove or comment out the author notes below if they aren't used.
%
% Primary Equal Contribution Note
%\Yinyang These authors contributed equally to this work.

% Additional Equal Contribution Note
% Also use this double-dagger symbol for special authorship notes, such as senior authorship.
%\ddag These authors also contributed equally to this work.

%\textcurrency a Insert current address of first author with an address update
% \textcurrency b Insert current address of second author with an address update
% \textcurrency c Insert current address of third author with an address update


% Group/Consortium Author Note
%\textpilcrow Membership list can be found in the Acknowledgments section.

% Use the asterisk to denote corresponding authorship and provide email address in note below.
* juyongp@kaist.ac.kr

\end{flushleft}

\justify

\section*{Abstract}
Online social media such as Twitter are widely used for mining public opinions and sentiments on various issues and topics.  The sheer volume of the data generated and the eager adoption by the online-savvy public are helping to raise the profile of online media as a convenient source of news and public opinions on social and political issues as well.  Due to the uncontrollable biases in the population who heavily use the media, however, it is often difficult to measure how accurately the online sphere reflects the offline world at large, undermining the usefulness of online media.  One way of identifying and overcoming the online-offline discrepancies is to apply a common analytical and modeling framework to comparable data sets from online and offline sources and cross-analyzing the patterns found therein.  In this paper we study the political spectra constructed from Twitter and from legislators' voting records as an example to demonstrate the potential limits of online media as the source for accurate public opinion mining.

% Please keep the Author Summary between 150 and 200 words
% Use first person. PLOS ONE authors please skip this step. 
% Author Summary not valid for PLOS ONE submissions.   
%\section*{Author Summary}
%Lorem ipsum dolor sit amet, consectetur adipiscing elit. Curabitur eget porta erat. Morbi consectetur est vel gravida pretium. Suspendisse ut dui eu ante cursus gravida non sed sem. Nullam sapien tellus, commodo id velit id, eleifend volutpat quam. Phasellus mauris velit, dapibus finibus elementum vel, pulvinar non tellus. Nunc pellentesque pretium diam, quis maximus dolor faucibus id. Nunc convallis sodales ante, ut ullamcorper est egestas vitae. Nam sit amet enim ultrices, ultrices elit pulvinar, volutpat risus.

%\linenumbers

\section*{Introduction}

The proliferation of online social media services such as Twitter is widely recognized as signifying a revolution
in how we utilize information and understand the way our society communicates.
It also presents an opportunity open and accessible to any interested party to utilize the massive data accrued from such services
for understanding the world as well as ourselves \cite{lazer_social_2009,giles_computational_2012}.
Among many attempts to harness the potential of online social media,
a prominent one is to mine the opinions and sentiments of the public for a variety of purposes, either academic or commercial
\cite{kwak_what_2010,cha_measuring_2010,golder_diurnal_2011,conover_political_2011}.
The use of social media for such purposes poses a potentially critical problem,
however: Since online social media users constitute an uncontrolled, thus likely unfair, sample of the population unlike in well-designed opinion polls,
it is unclear to what extent the online sphere accurately represents the offline world.
User profiles collected by online social media and made available to researchers frequently fall short of giving a sufficiently accurate demographic information about the users, and the traditional polling techniques using questionnaires have to be used to augment them \cite{duggan_socialmedia_2013}. The ideal case would be that the demographics of online users were identical to that of the offline public at large, guaranteeing the agreement of the opinions mined from the online sphere and with traditional polls. A less ideal, but a manageable case would be when the online demographics are precisely known, allowing us to employ straightforward calibrations to find accurate results.
But for a limited number of cases the demographic differences between online service users and non-users are not precisely known \cite{hargittai_whose_2007}, and given the breadth of topics discussed online and the difficulty of accurately estimating the number of users it does not seem realistic that such a state can ever be reached.

To those looking to mine for information from social media as a proxy for conducting polling,
the potential disparity between the online and the offline world due to the unknowable demographics raises
serious doubts on the usefulness of online social media, particularly where the offline reality is essential such as politics.  In such cases
the lack of a systematic method by which to compare the online and the offline would make finding relevant, useful information from online data infeasible. When researchers study such problems,
they could circumvent this issue by restricting the scope of their study to the online sphere so that by design the real-world public at large is of little concern.
This may be appropriate when the issue being studied is pertinent only to the online world; for instance,
when the issue affects the online sphere only, or when one declares that their only interest is in the opinions of the online sphere.
While certain scientific discoveries can still be made this way since the data themselves are nonetheless novel and interesting, the inability to draw conclusions regarding many issues pertinent to the offline potentially cripples the potential of online social media.

In order to overcome such a problem, it is important to identify the discrepancies between the online and the offline quantitatively. One way of achieving that is to apply a common analytical and modeling framework to comparable data sets representing the online and the offline so that we can perform a cross-analysis of the patterns found therein.  In this paper we demonstrate the process using as an example the landscape of political spectra (signifying the political leanings or positions) constructed from distinct online and offline data of the US and South Korea, two nations with modern representative democracies.  Specifically we study the political spectra of the legislators in each country, constructed in two ways. First, we construct a spectrum from the Twitter followership network, which we take as representing the online political landscape. There are some direct evidence that connection on social networks are biased towards those with similar political viewpoints. In a well-known study, Adamic~and~Glance showed a clear partisan divide on the blogosphere~\cite{adamic_political_2005}. Regarding Twitter, Parmalee~and~Bichard showed that over 70\% of those espousing strong conservative or liberal ideology completely avoided following political leaders who challenge their beliefs~\cite{parmelee_politics_2012}. In the case of Korea (one of the two countries studied in this work), Chang found that both politicians and ordinary Twitter users exhibit a systematic relationship between their political choices and their position in the Twitter network~\cite{chang_Twitter_2011}. Second, we construct a spectrum from the legislators' roll call (voting) records, which we take as representing the offline political landscape that more closely reflects the true public one, since the legislators' voting is  likely to be influenced by the offline public. The cross-examination of the two spectra should then reveal the differences between the online and the offline.

\begin{figure}[h!]
\centerline{\includegraphics[width=.9\textwidth]{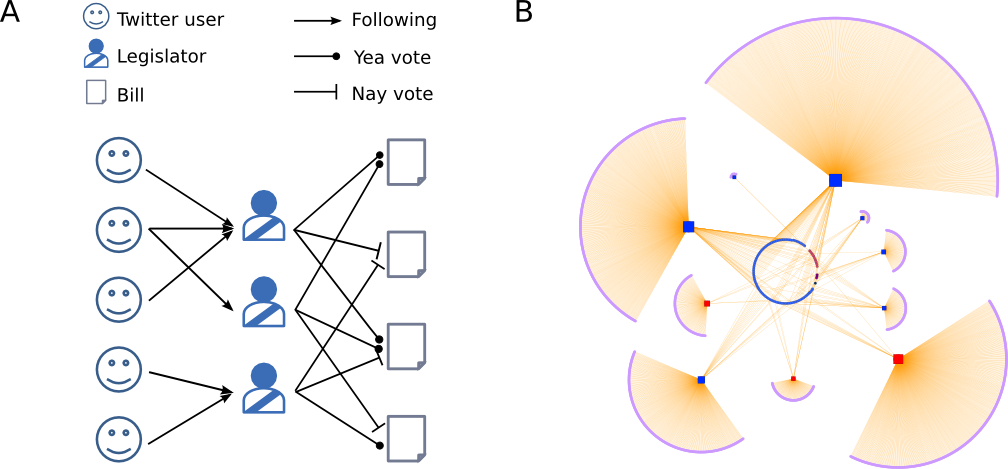}}
\caption{ 
(A) The schematic illustrations for the data employed in our study. The legislators are included in two distinct bipartite networks. On the left is the legislator--Twitter user network, and on the right the legislator--legislative bill network.  Of the two types of edges -- a `nay' vote and a `yea' vote -- we consider the `nay', since they are believed to carry more information in determining the legislators' political spectra (see the Method section for more detail).  The structure of each bipartite network can reveal differences in political positions of the legislators, which is the origin of the online-offline discrepancy. Here, for example, the upper two legislators occupy similar positions that are different from that of the the lower one since their follower sets are disjoint. The two groups' voting patterns may show less clear differences. In this study we use Multidimensional Scaling (MDS) and Kendall's ranking correlation coefficient to quantify the spectra and their discrepancies.
(B) A sample of the US senator--Twitter followership network consisting of ten legislators and their Twitter followers. The red squares are Republican (GOP) senators, and the blue squares are the Democratic (DEM) senators. All other nodes are Twitter users.
}\label{Fig:SchematicIllustration}
\end{figure}

\begin{figure}[h!]
\centerline{\includegraphics[width=.9\textwidth]{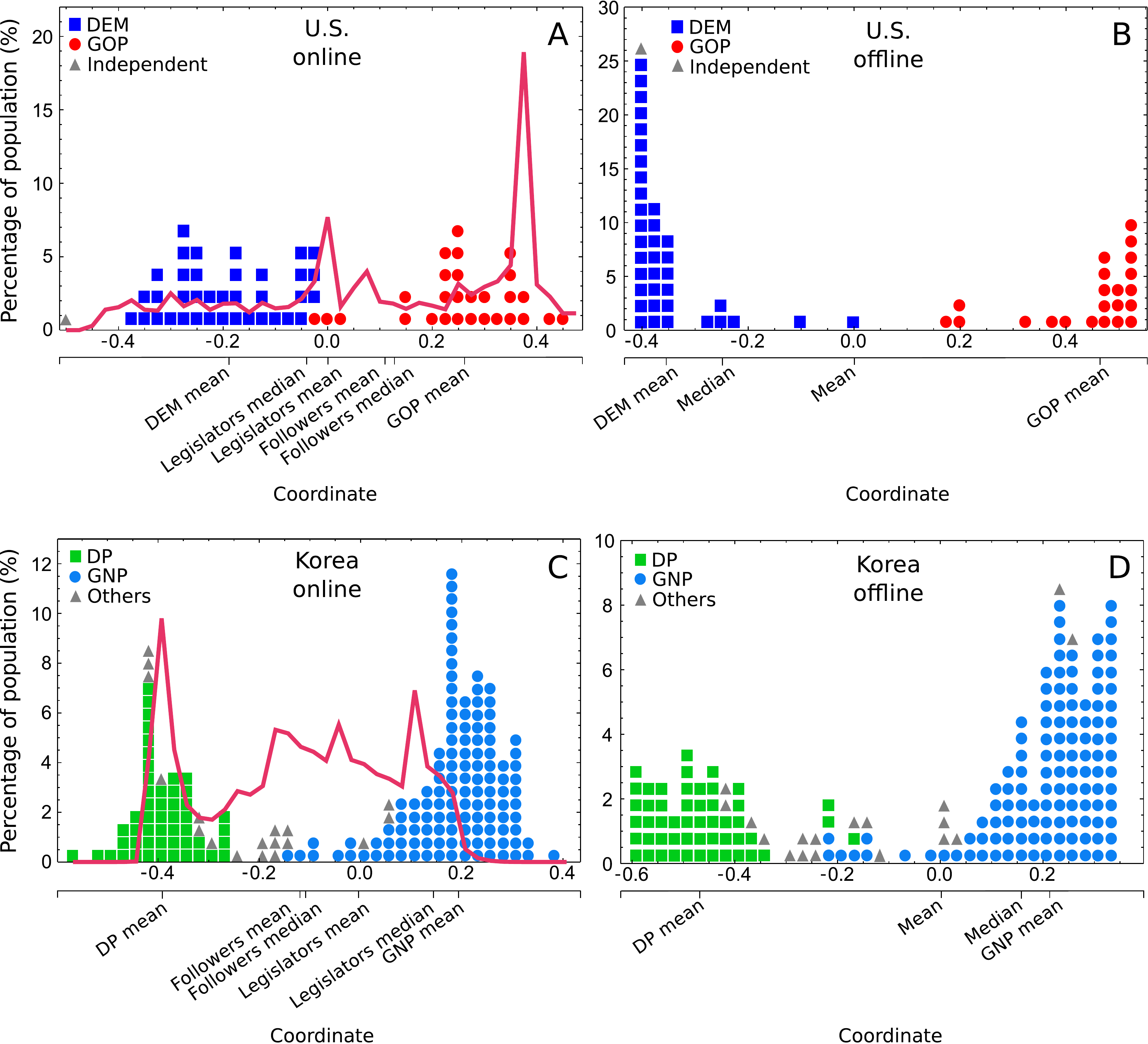}}
\caption{
The political spectra determined from Twitter and roll call data.
(A) The Twitter-based spectrum of the legislators, and the spectrum of Twitter users in the US;
(B) The roll call-based spectrum of the legislators in the US;
(C) The Twitter-based spectrum of the legislators, and the spectrum of Twitter users in Korea;
(D) The roll call-based spectrum of the legislators in South Korea.
The legislators are indicated by the shapes (one shape corresponds to one legislator) and the spectra of the Twitter users are indicated by solid lines.
The party-line splits of the legislators are evident in both the Twitter- and the roll call-based spectra for both countries.
The general Twitter public (solid lines in Figures A and C) show a majority-minority reversal, more people aligning themselves with the minority parties.
}\label{Fig:MDSResult}
\end{figure}

\section*{Results}

\subsection*{Data}

We utilized the following four data sets:
(i) The Twitter followership network of the members of the 111th U.S. Senate; (ii) the roll call of the Senators on legislative bills; (iii) The Twitter followership network of the members of the 18th Korean National Assembly; and (iv) the roll call of the Assembly members between the years 2008 to 2010 (covering three quarters of the term).  The Twitter data are current as of June 2011, and Twitter users who followed two or more legislators were included in the data.  The data contain all legislators who owned an official Twitter account at the time of data collection and maintained their seat during the entire term considered. This resulted in 67 senators (out of 100), 698 roll calls, and 139\,806 Twitter users for the US, and 194 assembly members (our of 299), 1\,119 roll calls, and 124\,341 Twitter users for Korea.  Both countries have two-party systems, the Republican Party (GOP) and the Democratic Party (DEM) that account for 99\% of the US Senate, and the Grand National Party (GNP) and the Korean Democratic Party (DP) that account for 86\% of the Korean National Assembly.
\cite{note_party_name}

The data were modeled as bipartite networks. See Figure 1 for the schematic illustrations of the data.

\subsection*{Online and offline political spectra}
The data introduced above can be modeled as bipartite networks on which the political spectra are constructed using matrix decomposition methods \cite{poole_polarization_1984,porter_network_2005}. In this study we use Classical Multidimensional Scaling (CMDS) \cite{cox_multidimensional_2001,brazill_factor_2002}. The method requires a dissimilarity function that act as the distance between two legislators. We use the Kulczynski dissimilarity \cite{hubalek_coefficients_1982}.  In essence, CMDS regards the dissimilarity values as the Euclidean distances between legislators in $1$-dimensional space and determines the coordinates of the legislators that most closely matches the given distances.  We then interpret the coordinates as the political positions of the legislators.  Once the spectra of the legislators are determined this way using the Twitter network, we can construct the spectra of the Twitter users by averaging the positions of the legislators that each user follows (see the Methods section for more details).

We consider the offline, roll call-based spectra as revealing the true political positions of the legislators as they results from their actual actions of voting.  Furthermore, we accept the roll call-based spectra as more accurately reflecting true political positions of the offline public and the society, as the legislators' actions are likely to be heavily influenced by public interests.

The political spectra of the legislators thus constructed are shown in Figures 2A and 2B for the U.S., and in Figures 2C and 2D for Korea.  The spectra of the Twitter users (the average of the positions of the legislators that they follow) are shown as the solid lines in Figures 2A and 2C.

A close examination of these spectra can reveals the disparities between the online and the offline. First, we find a stark indication of the significance of the disparity between Twitter and the offline in the majority-minority reversal in both countries: The majority of Twitter users align themselves with the minority parties. The mean and the median positions, and the overall distribution of the spectra are shown in Figure 2A for U.S. and Figure~2C for Korea.  We find that in the US, Republican Senators John McCain (the presidential candidate in 2008) and Jim DeMint (of the populist Tea Party movement) boasting a disproportionately large number of followers are primarily responsible for this behavior.  In Korea, the DP (green) overwhelms the GNP (blue) in the Twitter sphere despite being outnumbered by a nearly two-to-one margin in the number of seats in the Assembly.

Second, the comparison between the roll call-based spectra of the legislators as the proxy for the offline public at large and the spectra of the Twitter users also reveals a similar disparity. Although we cannot directly compare the online and offline spectra on the same plot because of the differences in scale, the comparison of relative biases in the distribution is still valid.
For the U.S., the spectrum of the Twitter users is heavily skewed towards GOP in sharp contrast to the DEMs who are favored in the spectrum of the offline public (Figures 2A and 2B). The trend is similar in the Korean spectra: the offline public favors GNP in contrast to the Twitter user who favor DP (Figures 2C and 2D).

Third, there also exists a discrepancy between the real political positions of the legislators and their positions constructed from the Twitter data.
The party-line split in the positions of the legislators (indicated as solid shapes in the figures) is immediately noticeable for both nations,
and accordingly the roll call-based spectra and the Twitter-based spectra of the legislators exhibit some degree of overall agreement (similarity) as measured
by Kendall's $\tau$-coefficient (1 means identity; 0 means independence), with $0.68 \pm 0.02$ for US and $0.48 \pm 0.01$ for Korea (see Methods section for more detail). When we consider each party separately, however, we find a weaker agreement -- for the US parties, the $\tau$ is $0.463 \pm 0.003$ for GOP and $0.291 \pm 0.022$ for DEM.
For the Korean parties that is $0.029 \pm 0.005$ for GNP and $0.062 \pm 0.051$ for DP.

This indicates that Korean Twitter population is more noticeably impervious to the political positions of the legislators, resulting in the very low level of intra-party correlations between the Twitter based spectrum and the roll call based spectrum of the legislators. In US the correlation is more sizable.

These disparities between the Twitter users and the offline public show the limits of Twitter as a source for fair and accurate opinion mining, especially pertaining to political or social issues. In fact, the majority-minority reversal shows us that Twitter may be functioning as a de facto ``alternative media.''

% It is not evident that this nature would sustain after the majority of real world politics is replaced in the future and this is an interesting question need to be studied.

\subsection*{Quantifying partisanship of twitter users}
The spread structures of the Twitter users' spectra in Figures 2A and 2C imply that Twitter users are biased (show partisanship) in choosing which legislators to follow.  We can quantify the strength of a user's bias by measuring the probability that a user follows certain parties, and comparing it with the random probability that a user becomes connected to a party if the connections were made at random. If a user is unbiased they must be identical, and deviations between the two probabilities signify the strength of the bias.

A widely-used measure for the discrepancy between two probability distributions is the Kullback-Leibler (KL) divergence $D$. It is non-negative, i.e. $D\ge 0$; $0$ when two distributions being compared are identical, and  positive otherwise.  To account for random fluctuations we calculate the $Z$-score of KL divergences (see Methods section for detail) $Z_D$ as a function of the number of politicians (degree) $k$ that a user follows, shown in Figure 3. $Z_D\gtrsim 1$ implies partisanship, while $Z_D\lesssim -1$ implies the opposite. The radius of the disc around each plot point is proportional to the the log of the number of users with the given degree $k$. The figure shows the widespread nature of partisanship on Twitter \cite{adamic_political_2005}, with the exception of high-degree Twitter users who are more balanced in following each party (see Methods section for more detail).

% The figure indicates most of the Twitter public have politically biased following pattern.
% The political spectra of the Twitter users we obtain stem from coordination of these biases.

\section*{Discussion}

In this paper we proposed a method to compare the political spectra of the online and offline public. We found significant discrepancies between the online and the offline, demanding caution when one tries to use online social media such as Twitter as source for fair and accurate opinion mining.  A quantitative measurement of political biases found widespread partisanship among Twitter users.

While our work shows the potential limits of the correctness of online social media-based findings, it is still meaningful that were able to obtain the political spectra of Twitter users based on their behavior observed objectively.  This enables us to overcome uncertainty in methods based on self-assessment of political positions \cite{greenwald_implicit_1995,burdein_experiments_2006,nosek_implitic_2010} caused by the lack of objective, common measure of political positions agreed upon by people.  In our method, however, a person's political position is determined from the objective behavior of all other people in the data, allowing us to construct global and objective political spectra.

\begin{figure}[h!]
\centerline{\includegraphics[width=.6\textwidth]{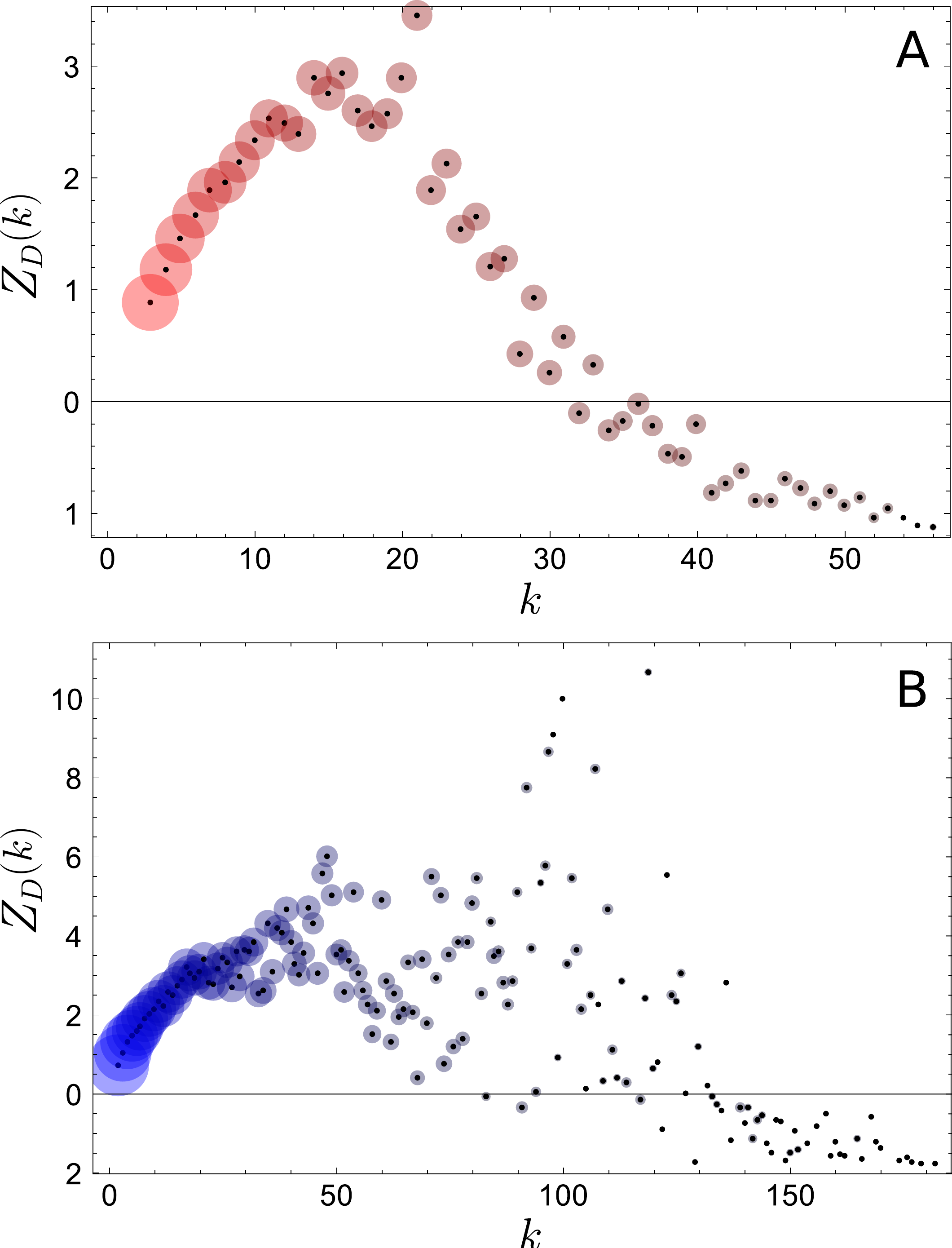}}
\caption{
The Z-score of the Kullback-Leibler (KL) divergence as a function of the number of legislators $k$ that a Twitter user follows (A) in the US, and (B) in Korea. The radius of each circle is proportional to the log of the number of Twitter users with given $k$. $Z_D\gtrsim 1$ implies a statistically significant level of partisanship (political bias). We see that partisanship is widespread in both countries, except for high-degree Twitter users for whom $D\to0$.
}\label{Fig:KLD}
\end{figure}

\begin{figure}[h!]
\centerline{\includegraphics[width=.6\textwidth]{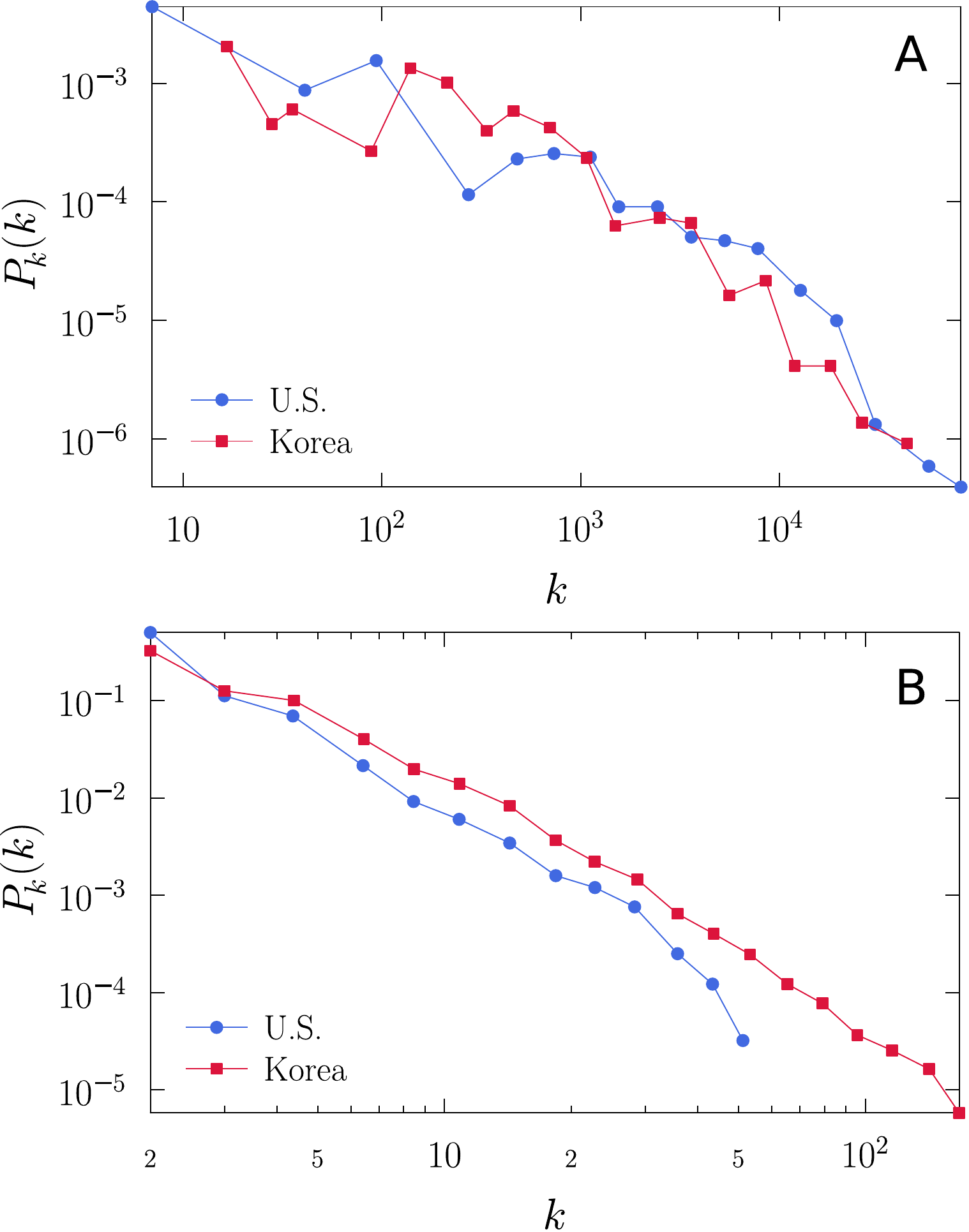}}
\caption{
(A) The distribution of the number of Twitter followers for each legislator.
(B) The distribution of the number of legislators that Twitter users follow.
All distribtuions are heavy-tailed.
}\label{Fig:TwitterDegreeDist}
\end{figure}

\section*{Methods}

\subsection*{Classical multi-Dimensional scaling}

CMDS is a method to embed a set of objects in a pre-defined $N$-dimensional space such that the distance between objects match given dissimilarities as much as possible. Specifically, CMDS aims to produce the coordinates of the objects that minimize a loss function called STRAIN \cite{cox_multidimensional_2001}.
In this paper we set $N=1$ to produce a linear political spectrum that agrees with the common way of thinking of political polarization (e.g. left wing versus right wing).

CMDS is performed by the eigendecomposition of the matrix made from the dissimilarities. The quality of the resulting coordinates in reproducing the given dissimilarities is defined as the fraction of the first eigenvalue over the sum of all positive eigenvalues. The values are 0.297 for Korea and 0.872 for US with the roll call data, and 0.113 for Korea and 0.093 for US with the Twitter data. The noticeably high value for the US roll call data may be due to the stable two-party system of the US.

While one could theoretically gain more information from higher-dimension CMDS, we could not observe meaningful patterns allowing straightforward political interpretation beyond the first. Our $1$-dimensional solutions show clear separations of the legislators along party lines, thus justifying the political interpretation given.

We used the Kulczynski dissimilarity \cite{hubalek_coefficients_1982}. In Twitter data it is given between two legislators $i$ and $j$ as
\begin{equation}
        \delta_T(i,j) = 1 - \frac{1}{2} \left| \frac{F_i \cap F_j}{|F_i|} + \frac{F_i \cap F_j}{|F_j|} \right|,
\end{equation}
where $F_x$  denotes the set of the followers of legislator $x$. The Kulczynski dissimilarity is recommended as an alternative to the commonly used Jaccard measure when the size of $F_x$ exhibits a wide range
as in our Twitter data (Figure 4) \cite{hausdorf_biotic_2003}.

The dissimilarity between legislators $i$ and $j$ in the roll call data is, similarly,
\begin{equation}
        \delta_R(i,j) = 1 - \frac{1}{2} \left| \frac{V_i \cap V_j}{|V_i|} + \frac{V_i \cap V_j}{|V_j|} \right|,
\end{equation}
where $V_x$ denotes the set of {\em nay} (opposition to a bill's passing) votes of legislator $x$. We ignore the {\em yea} votes here. The motivation for this is that most bills that pass are bipartisan, potentially limiting the discriminating power of the {\em yea} votes.

\subsection*{Notes on the spectrum of Twitter users}
From the legislator-Twitter data we estimated the spectrum of the legislators first on their Twitter followership, and then used it to obtain the spectrum of Twitter users.  A dual approach could have been employed where we perform CMDS on the Twitter users by defining the dissimilarities between the users based on the set of legislators that they follow, and determining the positions of the legislators based on the results. While there is no reason to object this approach in principle, its  computational cost was prohibitive, requiring several tens of gigabytes of memory to store the dissimilarity matrix alone.

\subsection*{Kendall's $\tau$ rank correlation coefficient}

Kendall's rank correlation $\tau$ quantifies the similarity between two orderings of objects.
Letting $(x_1,x_2,\cdots,x_n)$ and $(y_1,y_2,\cdots,y_n)$ be the orderings we wish to compare,
Kendall's $\tau$ rank correlation coefficient is defined as
\begin{equation}
\tau = \frac{2(C-D)}{n(n - 1)}.
\end{equation}
In this definition $C$ is the number of $(i,j)$ pairs such that the $(i,j)$ satisfies $x_i > x_j$ and $y_i > y_j$ or it satisfies $x_i < x_j$ and $y_i < y_j$.
$D$ is the number of $(i,j)$ pairs such that the $(i,j)$ satisfies $x_i > x_j$ and $y_i < y_j$ or it satisfies $x_i < x_j$ and $y_i > y_j$.

Errors in Kendall's $\tau$ were estimated through the jackknife method \cite{newman_monte_1999}.

\subsection*{Z-score of the Kullback-Leibler divergence}
Kullback-Leibler divergence (KL divergence) measures deviation of a probability distribution from a reference probability distribution. Here the reference probability distribution is the fractions of seats occupied by each party, i.e. the probability that a random user would follow a legislator from a given party.

Let $P_i$ be the fraction of the number of legislators from party $i$ among the number of the legislators, and
$p_{i j}$ be the fraction of the number of legislators from party $i$ followed by Twitter user $j$ among all legislators followed by $j$. Then the KL divergence $D_j$ of Twitter user $j$ is
\begin{equation}
D_j = \sum_{i=1}^N p_{i j} \log \frac{p_{i j}}{P_i},
\end{equation}
where $N$ is the number of parties. KL divergence is non-negative. A large $D_j$ means that user $j$'s followership pattern does not agree with the reference probability, i.e. favors one party disproportionately when in a two-party system.

While the average KL divergence $\left< D \right>_k$ of Twitter users who follow $k$ legislators can be obtained easily from the empirical data, for statistical significance we use the $Z$-score
\begin{equation}
Z_D(k) = \frac{\left< D \right>_k - \mu_k}{\sigma_k},
\end{equation}
where $\mu_k$ and $\sigma_k$ are the expectation and standard deviation of the KL divergence when a user choose to follow a party according to the reference probability $P_i$.
If we assume a Twitter user follows $k$ legislators randomly with uniform distribution,
then the probability of $\vec{n} = (n_1, \cdots, n_N)$, the vector of the number of legislators in each party that the user follows,
is given by the multinomial distribution $f(\vec{n}; \vec{P})$ where $n_1 + \cdots + n_N = k$ and $\vec{P} = (P_1, \cdots, P_N)$. Then we have
\begin{equation}
\mu_k = \sum_{n_1 + \cdots + n_N = k} f(\vec{n}; \vec{P}) \sum_{i = 1}^N \frac{n_i}{k} \log \frac{n_i}{k P_i}
\end{equation}
and
\begin{equation}
\sigma_k^2 = \sum_{n_1 + \cdots + n_N = k} f(\vec{n}; \vec{P}) \left( \sum_{i = 1}^N \frac{n_i}{k} \log \frac{n_i}{k P_i} \right)^2 - \mu_k^2.
\end{equation}

$Z_D(k)\gtrsim 1$ indicates that Twitter users follow a biased set of legislators, and that the bias is larger than the typical random fluctuation. Similarly, $Z_D(k)\lesssim -1$ means that Twitter users follow the parties in close agreement with the reference probability $P_i$. As $k$ approaches the actual number of legislators $p_{i,j}$ converges to $P_i$, $Z_D(k)$ becomes more negative (since $D\to 0$).

\section*{Supporting Information}

% Include only the SI item label in the subsection heading. Use the \nameref{label} command to cite SI items in the text.

\subsection*{S1 Dataset}
\label{SI:KRLegislatorList}
\href{http://journals.plos.org/plosone/article/asset?unique&id=info:doi/10.1371/journal.pone.0124722.s001}{\bf List of the Korean legislators.}
Each line is a record for a legislator.
The first column is the Twitter ID, and the second column is the name. The last column is the party of the legislator.
The file is encoded with UTF-8 for Korean characters.

\subsection*{S2 Dataset}
\label{SI:KRFollowershipData}
\href{http://journals.plos.org/plosone/article/asset?unique&id=info:doi/10.1371/journal.pone.0124722.s002}{\bf Followership data for the Korean legislators.}
Each line is a followee-follower relation.
The first column is the Twitter ID of a legislator, and the second column is the Twitter ID of a follower of the legislator.

\subsection*{S3 Dataset}
\label{SI:KRRollCallData}
\href{http://journals.plos.org/plosone/article/asset?unique&id=info:doi/10.1371/journal.pone.0124722.s003}{\bf Roll call votes data of the Korean legislators.}
Each line is a record of a legislator.
The first column is the name of the legislator, and the rest columns are votes.
A vote is encoded as 1 for yea, 2 for nay, 3 for abstention, and 4 for absence.
The file is encoded with UTF-8 for Korean characters.

\subsection*{S4 Dataset}
\label{SI:USLegislatorList}
\href{http://journals.plos.org/plosone/article/asset?unique&id=info:doi/10.1371/journal.pone.0124722.s004}{\bf List of the US legislators.}
The format is the same with \nameref{SI:KRLegislatorList}.

\subsection*{S5 Dataset}
\label{SI:USFollowershipData}
\href{http://journals.plos.org/plosone/article/asset?unique&id=info:doi/10.1371/journal.pone.0124722.s005}{\bf Followership data for the US legislators.}
The format is the same with \nameref{SI:KRFollowershipData}.

\subsection*{S6 Dataset}
\label{SI:USRollCallData}
\href{http://journals.plos.org/plosone/article/asset?unique&id=info:doi/10.1371/journal.pone.0124722.s006}{\bf Roll call votes data of the US legislators.}
The format is the same with \nameref{SI:KRRollCallData}.

\section*{Acknowledgments}
We thank Z. Hyung, C. Han, and B. Kahng.

%\nolinenumbers

%\section*{References}
% Either type in your references using
% \begin{thebibliography}{}
% Text
% \end{thebibliography}
%
% OR
%
% Compile your BiBTeX database using our plos2015.bst
% style file and paste the contents of your .bbl file
% here.
% 
%\begin{thebibliography}{10}
%\bibitem{bib1}
%Devaraju P, Gulati R, Antony PT, Mithun CB, Negi VS. Susceptibility to SLE in South Indian Tamils may be influenced by genetic selection pressure on TLR2 and TLR9 genes. Mol Immunol. 2014 Nov 22. pii: S0161-5890(14)00313-7. doi: 10.1016/j.molimm.2014.11.005
%
%\bibitem{bib2}
%Huynen MMTE, Martens P, Hilderlink HBM. The health impacts of globalisation: a conceptual framework. Global Health. 2005;1: 14. Available: http://www.globalizationandhealth.com/content/1/1/14.
%
%\end{thebibliography}

%\bibliography{References}{}
%\bibliographystyle{plos2015}

% This section is for figure legends only, do not include
% graphics in your manuscript file.
%
%\begin{figure}
%\caption{
%{\bf Bold the first sentence.}  Rest of figure caption.  
%}
%\label{Figure_label}
%\end{figure}

\end{document}